\documentclass[sigconf]{acmart}

\usepackage[font=small,labelfont=bf]{caption}
\usepackage{hyperref}
\usepackage{tikz}
\usepackage{amsmath}
\usepackage{filecontents}
\usepackage{footnote}
\usepackage{diagbox}
\usepackage{listings}
\usepackage{booktabs}
\usepackage{multirow}
\usepackage[ruled,vlined]{algorithm2e}
\usepackage{subfigure}
\usepackage{tabularx}
\usepackage{url}
\usepackage{algorithmic}
\usepackage{graphicx}
\usepackage{textcomp}
\usepackage{xcolor}
\usepackage{enumitem}
\usepackage{listings}
\usepackage[framemethod=tikz]{mdframed}
\usepackage{balance}
\usepackage{lipsum} 
\usepackage[htt]{hyphenat}
\usepackage{caption} 
\usepackage{siunitx}
\def\eg{\emph{e.g.,}\xspace}

\def\ie{\emph{i.e.,}\xspace}
\def\etal{\emph{et al.}\xspace}
\def\vs{\emph{vs.}\xspace}

\newcommand{\one}{({\em i}\/)}
\newcommand{\two}{({\em ii}\/)}

\AtBeginDocument{%
  \providecommand\BibTeX{{%
    \normalfont B\kern-0.5em{\scshape i\kern-0.25em b}\kern-0.8em\TeX}}}

\newcommand{\totalApps}[0]{\num{20195}~}

\newcommand{\familyApps}[0]{\num{3627}~}
\newcommand{\normalApps}[0]{\num{16568}~}
\newcommand{\appsWithRatings}[0]{\num{9453}~}

\newcommand{\numberComments}[0]{\num{13132577}~}
\newcommand{\appsWithComments}[0]{\num{11831}~}
\newcommand{\trackersNotAllowedFamilyPerc}[0]{81.25\%~}
\newcommand{\trackersNotAllowedFamilyPercExclude}[0]{13.08\%~}
\newcommand{\inconsistentAppsPerc}[0]{19.25\%~}
\newcommand{\inconsistentAppsHighPerc}[0]{9.99\%~}
\newcommand{\inconsistentAppsHighFamilyPerc}[0]{13.25\%~}
\newcommand{\inconsistentAppsHighNormalPerc}[0]{8.99\%~}
\newcommand{\fineLocationFamilyPerc}[0]{3.58\%~}

\newcommand{\bothLocationFamilyPerc}[0]{2.32\%~}
\newcommand{\eitherLocationFamilyPerc}[0]{4.47\%~}
\newcommand{\fineLocationNormalPerc}[0]{36.03\%~}

\newcommand{\bothLocationNormalPerc}[0]{28.31\%~}
\newcommand{\eitherLocationNormalPerc}[0]{39.83\%~}

\renewcommand\footnotetextcopyrightpermission[1]{}

\begin{document}

\title{Not Seen, Not Heard in the Digital World!\\ Measuring Privacy Practices in Children's Apps}

\author{
Ruoxi Sun$^\ast$$^\dagger$,  
Minhui Xue$^\dagger$$^\ddagger$,
Gareth Tyson$^\S$, 
Shuo Wang$^\dagger$$^\ddagger$,
Seyit Camtepe$^\dagger$$^\ddagger$,
Surya Nepal$^\dagger$$^\ddagger$}
\affiliation{\institution{$^\ast$The University of Adelaide, Australia}\country{}}
\affiliation{\institution{$^\dagger$CSIRO's Data61, Australia}\country{}}
\affiliation{\institution{$^\ddagger$ Cybersecurity CRC, Australia}\country{}}
\affiliation{\institution{$^\S$Hong Kong University of Science and Technology (GZ), China}\country{}}

\renewcommand \authors{Ruoxi Sun, Minhui Xue, Gareth Tyson, Shuo Wang, Seyit Camtepe, and Surya Nepal}

\renewcommand{\shortauthors}{R. Sun, M. Xue, G. Tyson, S. Wang, S. Campete, and S. Nepal}

\begin{abstract}
The digital age has brought a world of opportunity to children. Connectivity can be a game-changer for some of the world’s most marginalized children. 
However, while legislatures around the world have enacted regulations to protect children's online privacy, and app stores have instituted various protections, privacy in mobile apps remains a growing concern for parents and wider society. In this paper, we explore the potential privacy issues and threats that exist in these apps. 
We investigate \totalApps mobile apps from the Google Play store that are designed particularly for children (Family apps) or include children in their target user groups (Normal apps). 
Using both static and dynamic analysis, we find that \eitherLocationFamilyPerc of Family apps request location permissions, even though collecting location information from children is forbidden by the Play store, and \trackersNotAllowedFamilyPerc of Family apps use trackers (which are not allowed in children's apps). 
Even major developers with 40+ kids apps on the Play store use ad trackers. Furthermore, we find that most permission request notifications are not well designed for children, and \inconsistentAppsPerc apps have inconsistent content age ratings across the different protection authorities. 
Our findings suggest that, despite significant attention to children’s privacy, a large gap between regulatory provisions, app store policies, and actual development practices exist. Our research sheds light for government policymakers, app stores, and developers.
\end{abstract}

\pagestyle{plain}

\maketitle

\section{Introduction}

The last decades have seen a dramatic increase in our reliance on mobile services. 
Particularly due to COVID-19, more and more young people are spending time using mobile applications for entertainment, remote work, online learning, and day-to-day tasks~\cite{serra2021smartphone}. 
Unfortunately, the Internet is home to a vast amount of content, which is potentially harmful to children, including uncensored sexual imagery, violent content, and strong language~\cite{unicef2020child,barrie2017How}. 46\% of parents say their children, aged 11 or younger, who uses YouTube have encountered videos that were inappropriate for their age~\cite{Auxier2020Parenting}. Furthermore, children may expose sensitive data online, with more than 1/3 of young people in 30 countries reporting being a victim of online bullying~\cite{unicef2019More}.

Due to these concerns, many international jurisdictions have enacted privacy laws and regulations to promote and protect the privacy of children. 
These include the Children's Online Privacy Protection Act (COPPA)~\cite{coppa} and its implementation, the Children's Online Privacy Protection Rule (COPPR)~\cite{coppr}, which ``imposes certain requirements on operators of websites or online services directed to children under 13 years of age'', the European General Data Protection Regulation (GDPR)~\cite{EU:16}, and the Privacy Act 1988~\cite{PrivacyAct1988} in Australia. 
These regulations restrict the behavior of apps and define the obligations of app developers. 
For example, GDPR Art. 8, COPPA, and COPPR \S312 requires that applications obtain verifiable parental consent prior to collecting personal information from children. Apps must make reasonable efforts to ensure that the notification is received by a parent of the child. 
Although various privacy assertions are required in app stores (such as the permission list and the privacy policies), it is usually difficult for regular users to understand the potential threats an app may pose, let alone identify undesired or malicious application behaviors. 

To help parents determine age-appropriate mobile apps for their children, app stores have released strict developer policies, along with inspection and vetting procedures before app publishing. 
Critical app information is provided to help users understand the app before using it (such as the number of installs, requested permissions, a rating score, the name of developer, and the comments by other users).
Furthermore, every app that is sold through the Google Play store is rated for age-appropriateness. These content age rating systems recognize that mobile apps now run the entire gamut from interactive picture books for toddlers, through to graphic adult content. Hence, parents can use ratings to help with making app purchasing decisions. In 2015, Google Play launched the ``Designed for Families'' program~\cite{designedForFamilies}, which allows app publishers to opt into an additional review in order to have their apps labeled as being family-friendly (aiming at highlighting pre-approved, child-safe apps). 
Further, in 2020, Google Play added a ``Teacher Approved'' section, in which the Play store consults with teachers and specialists.
They rate these apps based on design, appeal, age appropriateness and the appropriateness of ads~\cite{teacherApproved}. However, considering that most information is provided by app developers' self-reporting (\eg by filling a form or answering questionnaires), a centralized age appropriateness rating system is still missing. 

We argue that the complexity of these different policies and systems challenge the ability of many parents (and children) to make informed decisions. 
To gain an understanding of this complexity, we provide the first comprehensive measurement study of privacy practices in children's mobile apps.
We inspect apps from both a technical and user-available information perspective, with the aim to expose improper children's app development practices and privacy threats. We particularly focus on apps that are 
\one~designed primarily for children under 13 and listed on the Children tab (Family apps, \ie can be found at \texttt{\url{https://play.google.com/store/apps/category/FAMILY}}); 
and 
\two~designed for everyone, \emph{including} children (Normal apps).
From the technical side, we conduct static and dynamic analysis of \totalApps apps, including \familyApps Family apps and \normalApps Normal apps.
From the users' perspective, we collect and analyze content age ratings from 5 different children's app rating authorities and \numberComments comments from \appsWithComments apps.
Our main findings are as follows:

\begin{itemize}[leftmargin=*]
\item Through static and dynamic analysis, we find that \eitherLocationFamilyPerc of Family apps request location permissions (which is forbidden by the Play store); 
\trackersNotAllowedFamilyPerc of Family apps use trackers that are not allowed to be used in children's apps. Even major players (having 40+ children's apps published in the Play store) do not follow the Play store policies.
    
\item We compare the content age ratings given by various agencies (these tag the suitability of apps for different age groups). We identify significant inconsistency among these different agencies. \inconsistentAppsPerc apps have such issues, with \inconsistentAppsHighFamilyPerc of Family apps and \inconsistentAppsHighNormalPerc of Normal apps having severe inconsistencies across authorities. We conclude that greater transparency should be introduced to help inform parents.

\item From users' comments, we find that Family app users complain more about app content, but less about privacy and security. Highly-rated apps have a larger number of complaints. We conclude that a more efficient mechanism to report and check inappropriate app content should be established.
    
\end{itemize}

Our findings suggest that despite significant attention paid to children's privacy by legislatures, app stores, and society, there is still a large gap between regulatory provisions, app store policies, and actual development practices.
We argue that app stores should establish more effective supervision mechanisms, provide more detailed information, reduce their reliance on developers' self-certified information, and respond more actively to user feedback. We believe our study can provide useful insights for government policymakers, app stores, developers, and researchers to build privacy-preserving apps for children. Our source code is publicly available at
\texttt{\url{https://github.com/children-privacy/children-privacy}}.

\noindent \textbf{Ethical considerations.~} 
All the apps and user comments in this research are collected from publicly available resources. The app and developer names mentioned in this paper are anonymized. All personal information (usernames, timestamps) is removed from the dataset. We have disclosed the findings to Google Play store and related entities. 

\section{Legal Background and Play Store Policies}

This section describes the legal background for online children's apps that are subject to regulations or policies, such as COPPA, GDPR, and the Google Play Store policies. We further briefly outline our legal analysis of potential violations in Android children's apps.

\noindent \textbf{Consent from parents.~}
Children merit specific protection with regard to their personal data, as they may be less aware of the risks related to the processing of personal data~\cite{gdprRecital38}. According to COPPR \S312.4~\cite{coppr}, 
apps with target users under 13 must make reasonable efforts, taking into account available technology, to ensure that a parent of a child receives \textit{direct notice} of the apps' practices with regard to the collection, use, or disclosure of personal information from children. 
As regulated in GDPR Art. 8, the processing of personal information is lawful only if the consent is given by the holder of parental responsibility over the child, meaning that users who are 15 years or younger need parental consent where applicable (member states can choose a younger age down to 13). 

Therefore, companies must obtain verifiable parental consent before gathering data from children below the age limit (13 years of age for COPPA, 16 for GDPR). This legal requirement implies that informing parents or legal tutors about data collection practices via the privacy policy is not sufficient, especially if the app disseminates sensitive data to third-party services. With this in mind, we wish to determine whether apps collect private data without user consent. As a result, any sensitive or personal data, particularly unique identifiers or geolocations, uploaded by the app to third parties without a user consent may be a violation of COPPA and GDPR. Note, in the Play store policy, apps that solely target children are not allowed to access location permissions. 

\noindent \textbf{Families policy requirements.~}
To better serve users, the Play store requires developers to provide accurate information about their apps. In addition to filling out the age rating questionnaire, developers also must provide details about their app's target audience and content. Depending on the target audience selections the developers make, the app will be subject to additional Google Play policies, to ensure that apps for children have appropriate content, show suitable ads, and handle personal and sensitive information correctly. 
For apps that are designed primarily for children under 13, they must participate in the Designed for Families~\cite{designedForFamilies} and comply with Google Play's Families Policy Requirements~\cite{familiesPolicyRequirements}. 
For any apps that have at least one target audience age group that includes children, developers must comply with Google Play's Families Policy Requirements. In short, for any apps that have target users that include children, Google Play's Families Policy Requirements are compulsory. Developers are responsible for ensuring their apps are appropriate for children and compliant with all relevant laws. Failure to satisfy the requirements may result in an app's removal or suspension. 

According to the Designing Apps for Children and Families policy by the Play Store~\cite{designedForFamilies}, apps designed specifically for children must participate in the Designed for Families program. This requires that apps can only use Google Play certified ad SDKs listed in Table~\ref{tab_dff_ad_sdks}.
% \gareth{Broken ref}
In order for an ad SDK to be included on the list, the ad SDK must self-certify that they are compliant with Play’s Families Ad Program policies and all applicable local laws and regulations. Apps that include children and adults in their target audience, but are not in the Designed for Families program, are allowed to use non-certified ad SDKs for serving ads only to users above the age of 13. Therefore, any non-certified ad SDK used in Family apps violates the Play Store policy.
Normal apps that have non-certified ad SDKs but do not implement a neutral age screen may also violate the policy.

\begin{table}[t]
\centering
\caption{Ad SDKs that participate in Play’s Families Ads Program}
\label{tab_dff_ad_sdks}
\resizebox{\linewidth}{!}{
\begin{tabular}{lll}
\toprule
\textbf{Ad SDKs} & \textbf{Code Signature} & \textbf{Network Signature} \\ \midrule

AdColony & \begin{tabular}[t]{@{}l@{}}com/adcolony/,\\ com/jirbo/adcolony/\end{tabular} & adcolony.com \\\midrule

% \rowcolor[HTML]{EFEFEF}
AppLovin & com/applovin & \begin{tabular}[t]{@{}l@{}}applovin.com,\\ applvn.com\end{tabular} \\\midrule

Chartboost & com/chartboost/sdk/ & chartboost.com \\\midrule

Google AdMob & \begin{tabular}[t]{@{}l@{}}com/google/ads/,\\ com/google/android/gms/ads/, \\ com/google/android/ads/,\\ com/google/unity/ads/, \\ com/google/android/gms/admob\end{tabular} & \begin{tabular}[t]{@{}l@{}}2mdn.net,\\ google.com, \\dmtry.com, \\ doubleclick.com, \\doubleclick.net, \\ mng-ads.com\end{tabular} \\\midrule

InMobi & \begin{tabular}[t]{@{}l@{}}com/inmobi,\\ in/inmobi/\end{tabular} & \begin{tabular}[t]{@{}l@{}}inmobi.com,\\ inmobicdn.net,\\ inmobi.cn\end{tabular} \\\midrule

ironSource & com/ironsource/ & ironsrc.co \\\midrule

Kidoz & com/kidoz/sdk & kidoz.net/kidoz-sdk \\\midrule

SuperAwesome & \begin{tabular}[t]{@{}l@{}}tv/superawesome/sdk,\\ tv/superawesome/lib/\end{tabular} & superawesome.com \\\midrule

Unity Ads & \begin{tabular}[t]{@{}l@{}}com/unity3d/services,\\ com/unity3d/ads\end{tabular} & unity3d.com \\\midrule

Vungle & \begin{tabular}[t]{@{}l@{}}com/vungle/publisher/,\\ com/vungle/warren/\end{tabular} & vungle.com \\ \bottomrule
\end{tabular}
}
\end{table}

% \begin{table*}[t]
% \centering
% \caption{Ad SDKs that participate in Play’s Families Ads Program}
% \label{tab_dff_ad_sdks}
% \resizebox{0.85\linewidth}{!}{
% \begin{tabular}{lll}
% \toprule
% \textbf{Ad SDKs} & \textbf{Code Signature} & \textbf{Network Signature} \\ \midrule
% AdColony & com/adcolony/, com/jirbo/adcolony/ & adcolony.com \\
% AppLovin & com/applovin & applovin.com, applvn.com \\
% Chartboost & com/chartboost/sdk/ & chartboost.com \\
% Google AdMob & \begin{tabular}[t]{@{}l@{}}com/google/ads/, com/google/android/gms/ads/, \\ com/google/android/ads/, com/google/unity/ads/, \\ com/google/android/gms/admob\end{tabular} & \begin{tabular}[t]{@{}l@{}}2mdn.net, google.com, dmtry.com, \\ doubleclick.com, doubleclick.net, \\ mng-ads.com\end{tabular} \\
% InMobi & com/inmobi, in/inmobi/ & inmobi.com, inmobicdn.net, inmobi.cn \\
% ironSource & com/ironsource/ & ironsrc.co \\
% Kidoz & com/kidoz/sdk & kidoz.net/kidoz-sdk \\
% SuperAwesome & tv/superawesome/sdk, tv/superawesome/lib/ & superawesome.com \\
% Unity Ads & com/unity3d/services, com/unity3d/ads & unity3d.com \\
% Vungle & com/vungle/publisher/, com/vungle/warren/ & vungle.com \\ \bottomrule
% \end{tabular}
% }
% \end{table*}

\noindent \textbf{Age ratings.~}
Age ratings (also known as maturity or content age ratings) rate the suitability of TV broadcasts, movies, comic books, or video games for its audience~\cite{ESRB,PEGI}.
This usually places a media source into one of a number of categories, to show which age group is suitable to view the media. In the Google Play store, the app developers are responsible for completing a rating questionnaire about the nature of the apps’ content. The ratings assigned to the app, displayed on Google Play, are determined by the questionnaire responses. Misrepresentation of an app’s content may result in removal or suspension~\cite{googleContentRating}

\begin{figure}[t]
    \centering
    \includegraphics[width=\linewidth]{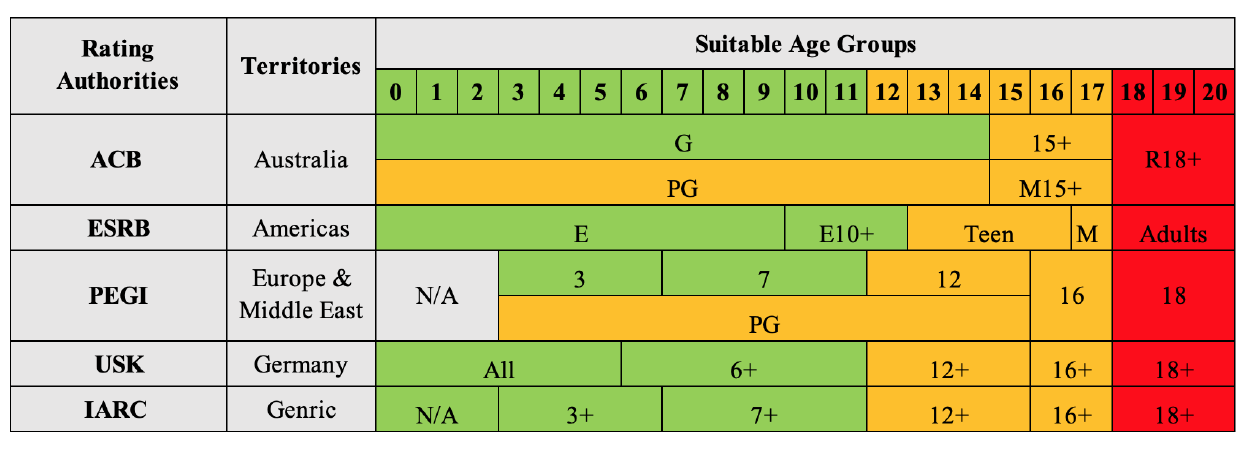}
    \caption{Age groups of content age ratings from different rating authorities.}
    \label{fig_rating_group}
\end{figure}

The ratings are intended to help consumers (especially parents) identify potentially objectionable content that exists within an app. Considering that, in different territories, rating standards can have differences and each rating authority uses its own methodology, an app can earn different ratings. 
In Figure~\ref{fig_rating_group}, we list a few rating authorities, including 
the Entertainment Software Rating Board (ESRB)~\cite{ESRB} in Americas, the Pan European Game Information (PEGI)~\cite{PEGI} in Europe and Middle East, Unterhaltungssoftware Selbstkontrolle (USK)~\cite{USK} in Germany, 
the Australian Classification Board (ACB)~\cite{ACB} in Australia, and the International Age Rating Coalition (IARC)~\cite{IARC}.
For some apps, their age ratings across different territories are inconsistent and confusing. 
For example, the app, \texttt{sg***ve}, earns a content age rating of ``PEGI 12'' in Australia, while in Germany, it is rated as ``USK 16+''. Note that, in the requirements of ``PEGI 12'', an app could contain ``slight violence towards fantasy characters'', ``non-graphic violence towards human-looking characters'', and ``mild bad language and no sexual expletives''. However, an ``USK 16+'' app is allowed to contain ``realistic violence'', ``shock and horror elements'', ``consistently explicit language'', and ``erotic or sexual focus'', which obviously cannot fall into the category ``PEGI 12'', even the gap between suitable age ranges of the two ratings is only 1 year (``PEGI 12'' is suitable for age group 12 to 15 and ``USK 16+'' is suitable for age group 16 to 17). 

\section{Analysis Pipeline}

\begin{figure*}[t]
\centering
\includegraphics[width=\linewidth]{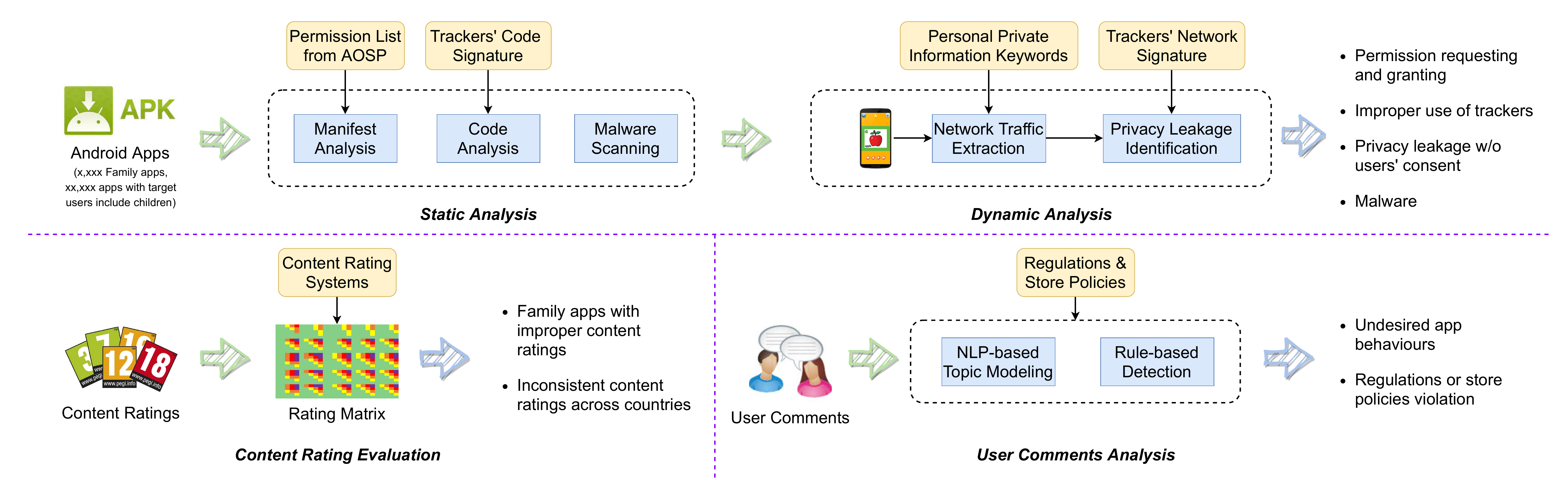}
\caption{An overview of the analysis pipeline}
\label{fig_overview}
\vspace{-3mm}
\end{figure*}

We present an overview of our analysis pipeline in Figure~\ref{fig_overview}. 
We combine the advantages of both static and dynamic analysis, as well as performing content (age) rating and user comment analysis. Our goal is to triage suspicious apps and analyze their behaviors in depth. 
Specifically, we evaluate apps (with targeted users including children) to find evidence of violations against privacy regulations and app store policies.
Our pipeline covers \totalApps different Android apps.
We check for improper use of trackers in children's apps, privacy leakage without consent, inconsistent age ratings from different authorities, and complaints by users. All of the apps are downloaded from the Google Play Store using a Google Play scraper. 

Our pipeline performs several tasks.
We first decompile the APK file of the Android app under-study, and conduct a static analysis to determine what permissions have been requested and what trackers are used in the app. Next, we execute each app version individually on a physical mobile phone with a network monitor, which allows us to observe apps’ run-time behaviors. We further collect the age ratings using different country settings and identify improper ratings according to inconsistent levels across rating agencies. Finally, we analyze users' comments with both NLP-based topic modeling and rule-based detection techniques to identify undesired app behaviors from users' complaints.

\subsection{Data Collection} \label{sec_data_collection}
We collect apps and their age ratings, as well as the user comments from Google Play store. Here we describe how we collect the data.

\noindent \textbf{App collection.~}
We wrote a Google Play Store scraper to download the most popular children's apps (English version) under each category. We collect \familyApps apps that participate in the Design for Families program.
We refer to this group as ``Family apps'' as their target users are purely children and families (categorized as ``Children'' in the Play Store).
We also collect \normalApps apps that are not particularly designed for children, but include children as target users.
We refer to these as ``Normal apps'' (with age ratings of ``Everyone'', ``Everyone 10+'', and ``Teen''). Note, we download all apps from the U.S. Google Play Store.
Because the popularity distribution of apps is long tailed, our analysis of the \totalApps most popular apps is likely to cover most of the apps that people currently use (from Mar 2021 to Dec 2021). We obtain executable APK files and corresponding metadata (\eg category and age rating) from Google Play store. 

\noindent \textbf{Age ratings collection.~}
We collect the age ratings of each app in the dataset from different countries by changing the country setting in the URL of each app when visiting the Play Store. 
We gather data for Australia, France, Germany, UK, and USA.
For example, the URL
\texttt{\url{https://play.google.com/store/apps/details?id=[package\_name]\&hl=en\&gl=US}} will lead us to the Play Store in the USA and if we change \texttt{gl=US} to \texttt{gl=FR}, it will present the app information to users from France. The reason we select these five countries is that they use different age rating standards that cover the main choices around the world, including ACB (Australia - games only), ESRB (North \& South America), PEGI (Europe \& Middle East), USK (Germany), and IARC (other countries). 

For a fair comparison, we only keep the age ratings of an app if it is available in all five countries. However, as we download all the apps from the  U.S. Play Store, there is no Normal App rated as ``Mature'' and ``Adults Only'' in ESRB in our dataset. Finally, we collect age ratings for \appsWithRatings apps.

\noindent \textbf{User comments collection.~}
We collect the \num{3000} most recent comments for each app. Note, as we aim to find undesired app behaviors and policy violations from user comments, we only collect the comments with rating stars below 2. This focuses our analysis on negative feedback. We further remove comments with fewer than 5 words. In total, we collect \numberComments comments for \appsWithComments apps. 

\subsection{Static Analysis}

Our static analysis focuses on three parts: Manifest Analysis, Code Analysis, and manual inspection.
We use these to measure the use of dangerous and signature permissions, as well as trackers in children's apps. 
To perform static analysis on the Android Package (APK) binary files, we first decompile the APK of each app to its corresponding \textit{class} and \textit{xml} files using AndroGuard~\cite{androGuard}.  
The de-compiled \texttt{AndroidManifest.xml} file is parsed to extract the permissions requested by the app. We then compare the Family apps and Normal apps with respect to the requisition of location related permissions (\ie \texttt{ACCESS\_COARSE\_LOCATION} and \texttt{ACCESS\_FINE\_LOCATION}).

In the tracker analysis, we extract all the class names in the de-compiled source code and match them with trackers' code signatures.
We adopt a list of trackers from Exodus Privacy with 693 known trackers~\cite{trackerlist} and extend the list with 32 more trackers reported in other online resources.
We then search for class names that contain any code signatures from the tracker list. For example, an app that contains \texttt{com/adcolony} in its source code will be detected as containing tracker ``AdColony'', which matches its signature \texttt{com/adcolony/, com/jirbo/adcolony/}. 
 
Note, our static approach may report false positives. It cannot determine whether the trackers are active or not during run-time. However, we argue that such static analysis still reflects potential tracking of users, especially when an app contains a large number of trackers. We further compare the use of trackers in Family and Normal apps. Finally, we manually review the Normal apps with most trackers to check whether there are age screens implemented.

\subsection{Dynamic Analysis}

Our static analysis focuses on detecting embedded third-party SDKs that potentially collect and disseminate personal children data to the Internet.
We compliment this with dynamic analysis, to collect evidence of personal data dissemination.  

\noindent \textbf{App execution.~}
To analyze whether personally sensitive information is leaked to third-parties without user consent, we rely on the automatic method previously proposed by Feal \etal~\cite{feal2020angel}.
We launch each app and run it for 5 minutes without interacting with it. This implies that we do not actively consent to data collection and we do not carry out any of the children actions, opting instead to leave the app running with no input. 

For this, we implement the dynamic analysis described in Figure~\ref{fig_overview}, which consists of 4 Xiaomi Notebook and 9 Android phones running a rooted Android 10.
This allows us to monitor the network traffic of each of the \totalApps Android apps. We automatically run each app using the Android Automator Monkey~\cite{monkey} without human intervention. Monkey is a UI fuzzer which simulates user input events, such as clicks, touches, or gestures, into the app. Concretely, we start each app with the Android Debug Bridge (ADB) command, \texttt{adb shell monkey -p [package name] n}, where the ``\texttt{-p}'' parameter specifies the package to run, and \texttt{n} indicates the number of events. Here, we set \texttt{n} as 1. 
After 5 minutes, we force stop and uninstall the app.
The logs are then cleared and the device is ready to be used for the next test. We store the resulting network traffic in a database for offline analysis, which we discuss later. 

\noindent \textbf{Network monitoring.~}
We monitor all network traffic, including TLS-secured flows, using a network monitoring tool, Lumen Privacy Monitor~\cite{razaghpanah2018apps}.
This has shown to be effective in several prior research activities~\cite{han2020price,feal2020angel,gamba2020analysis,reardon201950}. The network monitoring module leverages Android’s VPN API to redirect all the device’s network traffic through a localhost service that inspects the traffic, regardless of the protocol used. The network streams are reconstructed and linked back to the original app through mapping the UID obtained from \texttt{proc} filesystem to the socket owned by app.
In addition, TLS interception is also enabled through the system trusted root certificate installed into the system, which allows Lumen to decrypt TLS network traffic~\cite{razaghpanah2018apps}. Note that, due to the limitation of dynamic testing (\ie we may miss catching network traffics), our findings can only provide a lower bound of traffic monitoring.

\noindent \textbf{Personal information in network flows.~}
We define ``personal information'' as any piece of data that relates to an identified or identifiable individual or data subject and could distinguish them from another.
It is well known that third-party advertisement networks, online service providers, and mobile app developers track users' personal information across kinds of interfaces, such as devices, websites, and mobile apps, to better target ads or potential customers.
For this reason, we look into apps’ access to the persistent identifiers that enable long-term tracking, such as the geolocation information.

We focus on detecting apps that access specific types of sensitive data without user consent. Notably, the unauthorized collection of geolocation information in Android has been the subject of prior regulatory action~\cite{familiesPolicyRequirements}. 
In Table~\ref{tab_pii}, we list the summary of personal identifiable information that is checked in the network traffic data, as well as examples of corresponding keywords/values we search for in the traffic. Note, some of the PII (\eg device model and brand) may be needed for proper app functionality and is generally not considered sensitive within the industry. Therefore, we consider them as low risk.

\begin{table}[t]
\centering
\caption{Summary of personal identifiable information checked in network transmission data}
\label{tab_pii}
\resizebox{\linewidth}{!}{
\begin{tabular}{p{1.2cm}p{4.5cm}p{2.5cm}l}
\toprule
\textbf{PII}& \textbf{Description} & \textbf{Example Keywords}& \textbf{Risk}  \\ \midrule
Device Model & Identifies device model and manufacturer. & Redmi Note9 Pro & Low \\ \midrule
Brand & Identifies phone brand. When combined with other information, it can be used to identify a user uniquely. & xiaomi & Low \\ \midrule
Board Info & Identifies hardware and the phone model. & miatoll & Low  \\ \midrule
Build number & Identifies uniquely the Android OS and the version. & QQ3A.200905.001 & Low  \\ \midrule
MAC Address & Identifies uniquely the WiFi AP users are connecting to, leaking users' activities and location. &  & Mid  \\ \midrule
Private IP & By leaking the private IP address of your device, an ad network, tracker or application developer can better identify unique users. & 129.127.146.*** & Mid  \\ \midrule
Device Fingerprint & A fingerprint of user device which can be used by analytics and ad services to track user. & google/ walleye/ walleye:8.1.0
OPM1.171019.011/ 4448085:user/ release-keys & High  \\ \midrule
Location & The location data of a user. & "\$country", "\$city" & High  \\ \midrule
Timezone & Identifies the current timezone. & America/New\_York & High  \\ \midrule
IMEI & The IMEI (International Mobile Station Equipment Identity) identifies the device uniquely,  which could be used to track user's traffic and online behavior. & 866400053132507 & High  \\ \midrule
Serial number & Allows ad networks and online trackers to identify a user uniquely for tracking, surveillance or advertising purposes. & 3a9eb795 & High  \\ \midrule
Advertising ID & A unique string of characters that identifies the user’s device, for purposes like measuring app usage and ad personalization. & 7cba4b19-3ee3-4c14-9ec8-10ca1ad1abe1 & High  \\ \bottomrule
\end{tabular}
}
\end{table}

\subsection{Content Age Rating Evaluation}
\begin{figure}[t]
    \centering
    \includegraphics[width=0.95\linewidth]{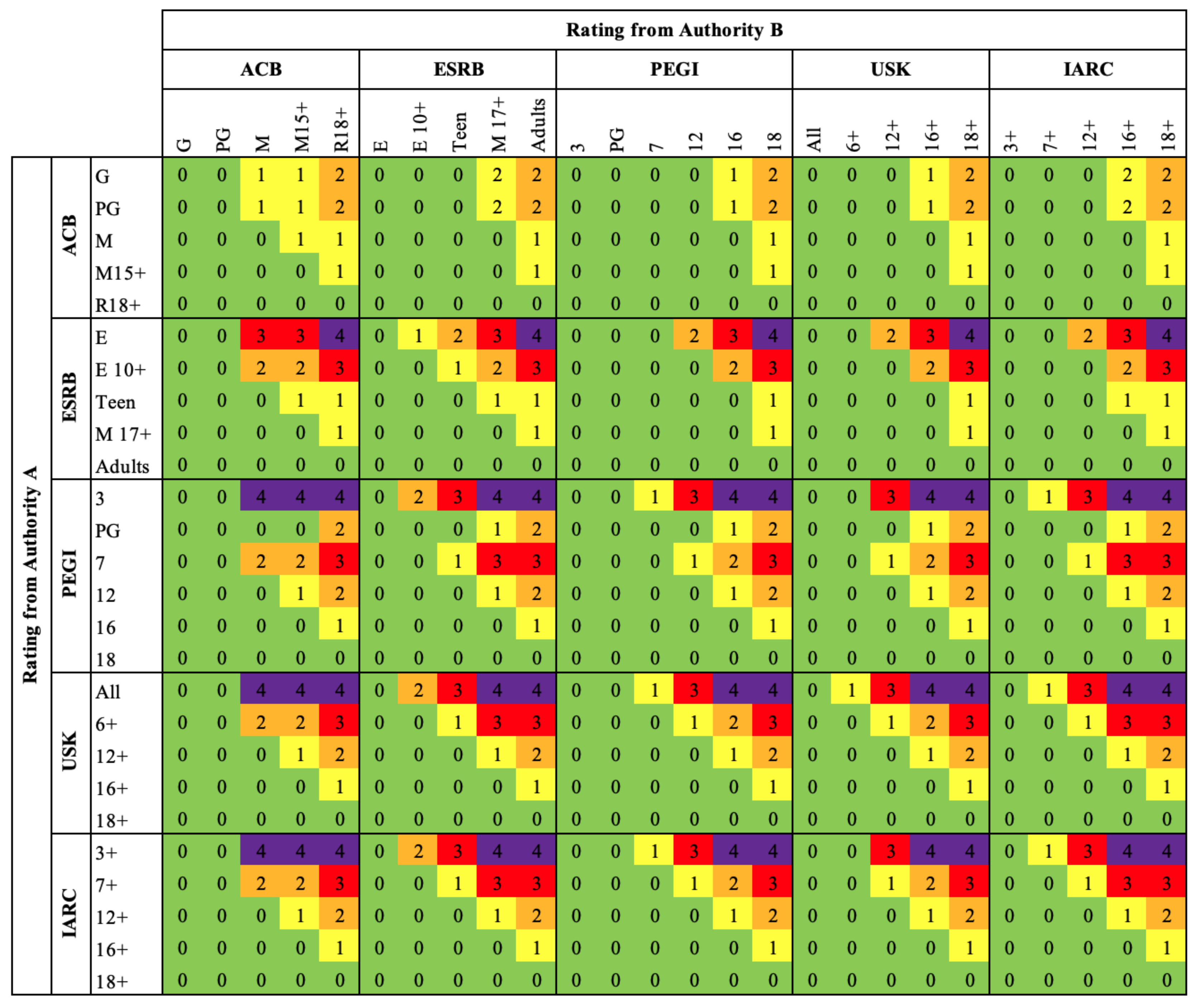}
    \caption{Inconsistency level matrix for age rating evaluation}
    \label{fig_rating_matrix}
\end{figure}

\noindent \textbf{Measuring rating inconsistency.~}
To investigate inconsistent age ratings, we measure the inconsistency with levels from 0 to 4, depending on the distance of suitable age groups between each category. 
Specifically, we consider the inconsistency between two ratings from authorities $A$ and $B$ in the following manner.
\one~We define the suitable age group for rating $A$ within range $[a_1,a_2]$, where $a_1$ is the minimal allowed age of rating $A$ and $a_2$ is the minimal allowed age of the next rating level minus 1. For example, the suitable age group for ``USK 6+'' is $[6,11]$, as the next rating level is ``USK 12+'' for which the minimal allowed age is 12. 
\two~We determine the inconsistency level between two ratings $A$ and $B$, according to the gap between two age groups, \ie $b_1-a_2$, as shown in Equation~\ref{equ_inconsistent_level}.

\begin{equation}\label{equ_inconsistent_level}
\small
Inconsistency~Level = \left\{ {\begin{array}{*{20}{l}}
{0,~~if~b_1-a_2 \leq 0},\\
{\lfloor\frac{(b_1-a_2)}{T}\rfloor+1,~~if~9 \geq b_1-a_2 > 0},\\
{4,~~otherwise.}
\end{array}} \right.
\end{equation}

\noindent
where $T$ is the threshold between two inconsistency levels (we increase the level by 1 for each $T$ years gap). We set $T$ as 3 as for most content age ratings the interval between neighboring levels is around 3. For example, if the two age groups overlap (\ie $b_1-a_2<0$), we determine the inconsistency level as 0; while for a 9 years gap, we determine the inconsistent level as 4.
To summarize the above, we present a matrix in Figure~\ref{fig_rating_matrix}, showing the level assigned to each combination of ratings. 
Using this, we later compare the age ratings of each app. 
Further, for any that has inconsistency levels larger than 3, we manually review them (including their description and Android packages provided across different territories).

\subsection{User Comments Analysis}

\begin{figure}[t]
\centering
\includegraphics[width=\linewidth]{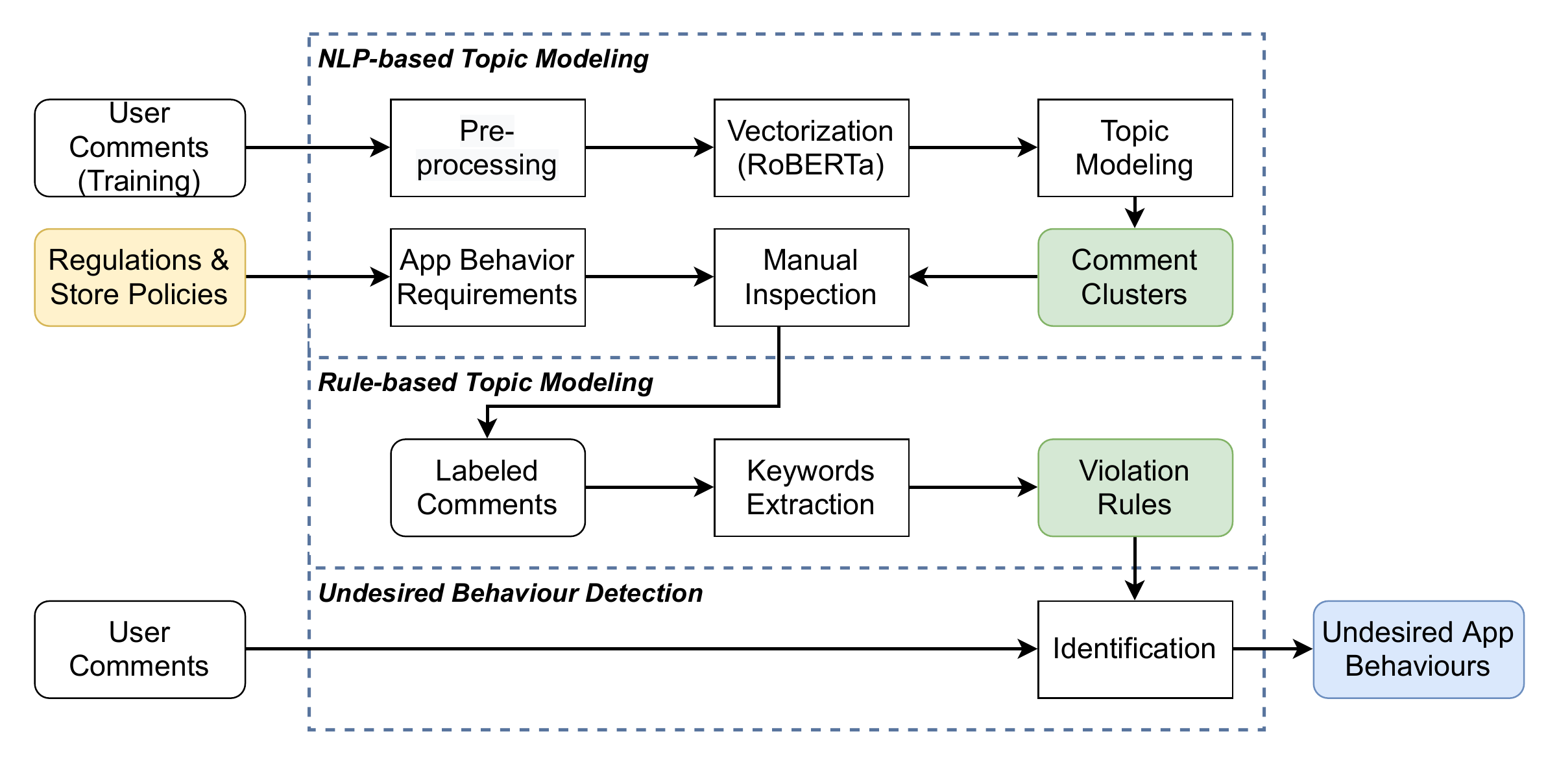}
\caption{Summary of user comment analysis pipeline}
\label{fig_comments_pipeline}
\vspace{-5mm}
\end{figure}

Beyond the static and dynamic analysis of apps, the comments left by users may also highlight the violation of regulations or policies, including user concerns. Recent research~\cite{hu2021champ} relies on user comments to identify violations against mobile application market policies, using a semi-automated rule-based process. 
Figure~\ref{fig_comments_pipeline} presents an overview of our comment analysis pipeline. 
We adopt natural language processing (NLP) techniques to interpret the comments and train machine learning models to categorize and identify informative comments. 
This comment analysis complements the static and dynamic analysis results (since some application behaviors may not be identified by technical analysis, but could be better identified by users). The rules will be further used in the recognition of regulation violations or undesired app behaviors. 

\noindent \textbf{Pre-processing and embedding.~}
We filter out comments with fewer than 5 words. We further translate emojis into words using the \texttt{emoji} Python package~\cite{emojiPackage}.
We utilize the pre-trained RoBERTa~\cite{liu2019roberta} embedding to vectorize each comment into a 300-dimensional vector. RoBERTa (Robustly optimized BERT approach) is a retraining of BERT~\cite{devlin2018bert} with an improved training methodology and 10x more training data to improve the BERT performance.
It achieves state-of-the-art results (2\% to 20\% improvement over BERT)~\cite{singh2021evolving}. 

\noindent \textbf{Comment clustering.~}
The next step of user comment analysis is to cluster the comments to tease out the different concerns that users describe.
These include complaints about functionality, performance, advertisement, personal data collection, vulgar content, violence, and payment deception. We refer to these as \textit{undesired behavior topics}. As the Play Store user comments do not have fine-grained labels for these topics, we use unsupervised learning to cluster comments into types.

We rely on $k$-means clustering using Cosine distance to identify the user concerns.
We use $k$-means clustering as our experiments expose good results (we leave exploration of other clustering solutions to future work). 
Without knowing how many distinct topics users may write about, the challenge of applying $k$-means in our clustering task is how to determine a proper $k$ value (\ie the number of target clusters). Recent work by Nema~\etal~\cite{nema2022analyzing} proposes a Summarization Metric as shown in Equation~\ref{equ_summarization_metric}, aiming for a $k$ that results in well separated clusters and a high number of compact clusters. 

\begin{equation}\small
\begin{split}
Summarization~Metric = dist_k * M_k,
\end{split}
\label{equ_summarization_metric}
\end{equation}

\noindent
where $dist_k$ is the minimum of the cosine distance between all pairs of $k$ cluster centers, and $M_k$ is the number of compact clusters. A compact cluster is a cluster in which at least 30\% of the samples have silhouette scores higher than the average silhouette score in this cluster.
We iterate through $k = 5, 10, 15, ..., 100$ and chose $k$ for which the summarization score is highest, as we want to increase both $dist_k$ (distance between cluster centers) and $M_k$ (number of compact clusters).
A larger $k$ may include more undesired behavior topics in the clustering results, but could also lead to a too heavy workload in the later manual inspection. Finally, we select $k$ as 40.

To summarize the topic of each cluster, we further conduct a manual inspection on the clustering results. Specifically, we select the top 20 (per cluster) representative comments that are nearest to the corresponding cluster centers and can be considered as representative of the entire cluster. 
We rank a comment's representativity using their silhouette scores. We carefully analyze these representatives manually across all clusters and determine the topic of each cluster. Considering that not all user comments focus on expression of concerns or complaints, we only keep the 18 clusters with topics related to undesired app behaviors, and drop the remaining 22 clusters. 

\noindent \textbf{Semantic rules extraction.~}
If we directly use $k$-means to recognize users' concerns, it may lead to a high false positive rate. 
This is because we can only classify each comment based on its distances to the cluster centers, and we have to assign a label even if it may not be relevant. Therefore, based on the clustering results, we propose a rule-based approach that assigns topics to comments according to rules, \ie if a comment is matched to a semantic rule, the comment will be recognized as belonging to the corresponding topic. Specifically, for each cluster, we first remove stopwords (according to the NLTK English stopwords list~\cite{stopwords}) and sort the remaining words in descending order based on TF-IDF weighting~\cite{rajaraman2011mining} to generate a word list. We then manually select keywords for each word list and merge the word lists that usually appear in the same topics. We finally obtain 15 topics in 4 categories as shown in Table~\ref{tab_topic}.

\begin{table}[t]
\caption{List of user comment topics.}\label{tab_topic}
\begin{resizebox}{\linewidth}{!}{
\begin{tabular}{@{}lp{8cm}@{}}
\toprule
\textbf{Categories} & \textbf{Topic Descriptions} \\ \midrule
\textbf{Content} & \begin{tabular}[c]{@{}p{8cm}@{}}
\textbullet~Containing violence, blood or scaring contents that are not proper for kids.\\
\textbullet~Containing sexual content that is not allowed for kids.\\
\textbullet~The app content encourages the use of tobacco or drugs.\\
\textbullet~Exposing inappropriate language to children.\\
\textbullet~Presenting depicting criminal activities to kids.\end{tabular} \\ \midrule
\textbf{Ads} & \begin{tabular}[c]{@{}p{8cm}@{}}
\textbullet~Providing too many advertisements.\\
\textbullet~The users are disrupted by ads.\\
\textbullet~Ads shortcuts in launching menu or notification bar.\\
\textbullet~Redirection or drive-by download by ads.\end{tabular} \\
\midrule
\textbf{Privacy} & \begin{tabular}[c]{@{}p{8cm}@{}}
\textbullet~Leaking or stealing users'   private information.\\
\textbullet~Abusing the permissions (e.g., requesting unnecessary permissions).\\
\textbullet~Collecting unnecessary private data.\\
\textbullet~Sharing data with third-parties or other users without user's consent.\end{tabular} \\
\midrule
\textbf{Security} & \begin{tabular}[c]{@{}p{8cm}@{}}
\textbullet~Containing virus or malware.\\
\textbullet~Being suspicious to payment fraud.\end{tabular} \\ \bottomrule
\end{tabular}
}
\end{resizebox}
\end{table}

For each topic, we obtain a keyword set containing one or more representative keywords.
Finally, we generate semantic rules in a form of $\{w_1, w_2, d, t\}$ for each keyword set, where $w_1$ and $w_2$ are two keywords, $d$ is the distance constrain between $w_1$ and $w_2$, $t$ is the topic that a matched comment will be assigned to.
Concretely, we first manually select 50 representative comments from each topic as labeled samples. Then we traverse each keyword set by calculating F1-scores for each single keyword (set $w_2$ as \texttt{null} and $d$ as 0) and each pair of keywords under different distance constraints (from 1 to 20), and select the rules with F1-score larger than 0.8. 
For example, a semantic rule could be $\{kid, improper, 2, not\_proper\_for\_kids\}$, which detects any comments that contains the keywords \texttt{kid} and \texttt{improper}, and the distance between the two keywords are smaller than 2 words.
Manual inspections are further conducted on a pilot set of \num{5000} labeled comments and rules with an error rate larger than 10\% are removed. 
Finally, we obtained 19 rules for user comment categorization.

We note that, as the semantic rules are extracted from representative 
comments in clusters, this approach may still focus more on the representative comments and can only provide a lower bound of the detection of users' complaints on undesired app behaviors or regulation violations. 

\section{Results}\label{sec_results}

In this section, we present the measurement results from applying our analysis pipeline to \familyApps children's apps that participate in the Google Family project (the ``Family Apps''), and \normalApps normal apps with kids included in the target user group (the ``Normal Apps''). We conduct a regression test and find that 2,857 (77.8\%) of detected apps are still listed in Google Play (for more than 1 year since we collected them). 

\subsection{The usage of Location Permissions}

Through the static analysis, we measure the use of location related permissions, including \texttt{ACCESS\_COARSE\_LOCATION} and \texttt{ACCESS\_FINE\_LOCATION} permissions, in the Family and Normal apps. 
We find that 162 (\eitherLocationFamilyPerc) Family apps request at least one of the location related permissions, asking for the access to user's coarse or fine location information (\fineLocationFamilyPerc for \texttt{ACCESS\_FINE\_LOCATION} and \bothLocationFamilyPerc for both), which is potentially violating the Design for Family policy. 
We further report the permissions usage in normal apps primarily as a comparison point to show that children's apps are requesting fewer permissions.
\eitherLocationNormalPerc of Normal apps request location permissions (\fineLocationNormalPerc for \texttt{ACCESS\_FINE\_LOCATION} and \bothLocationNormalPerc for both). 
However, note that all the Normal apps in our research have target users that include children (less than 13 years old). We argue that developers should be equally cautious about using location permissions when the target users include children.

\begin{figure}[t]
    \centering
    \includegraphics[width=0.9\linewidth]{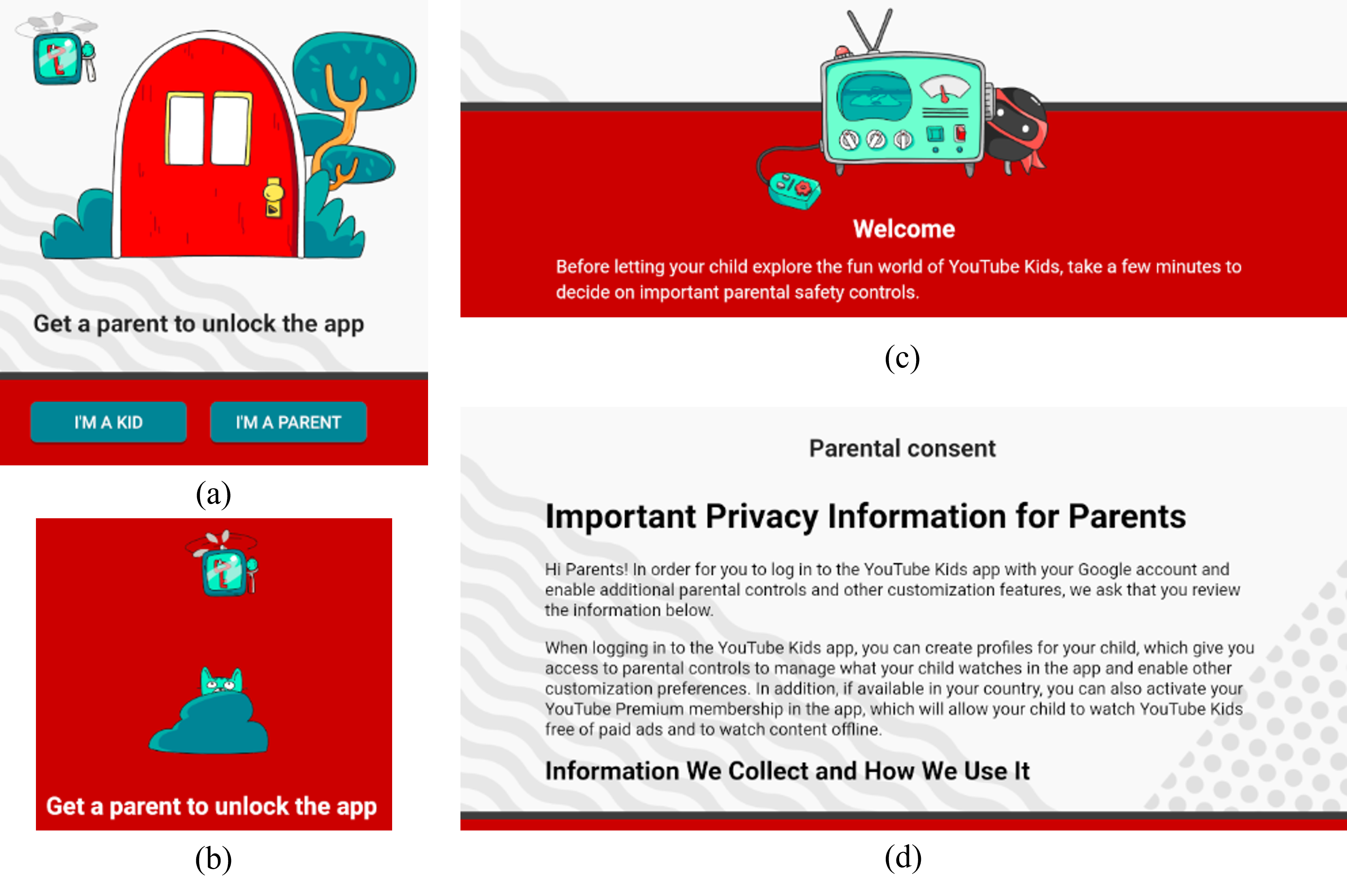}
    \caption{Examples of YouTube Kids notifications}
    \label{fig_youtube}
\end{figure}

\noindent \textbf{Case study: YouTube Kids.~}\label{youtubekids}
The YouTube Kids app~\cite{youtubekids}, a Google product built with children in-mind, aims to make it safer and easier for children to explore videos.
This provides a good example of permission request notifications for children. As shown in Figure~\ref{fig_youtube}(a), when the app is opened for the first time, a simple and clear notification is presented to users, noting that a parent is required to unlock the app. 
If the user selects ``I'M A KID'', the app requires the child to get their parent, and the only choice is going back to the previous screen (Figure~\ref{fig_youtube}(b)). Only after the ``I'M A PARENT'' button has been clicked, the app starts showing a introduction video (Figure~\ref{fig_youtube}(c)) and obtaining permissions from parent user (Figure~\ref{fig_youtube}(d)). Although a child may fool this process and login with their parent's account, the app tries its best to ask consent from guardians. Through our manual checks on 200 randomly selected Family and Normal apps, we did not find other permission notification designed for children users in other apps.

\vspace{2mm}
\begin{mdframed}[backgroundcolor=white!10,rightline=true,leftline=true,topline=true,bottomline=true,roundcorner=2mm,everyline=true,nobreak=false]  
\noindent\textbf{Finding summary:}  
\begin{itemize} [leftmargin=*]
\item \eitherLocationFamilyPerc of Family apps request location-related permissions, which is potentially violating the Design for Family policy. 
\item \fineLocationNormalPerc of Normal apps request the access of fine location information, where children users are not particularly unprotected.
\end{itemize} 
\end{mdframed}
\vspace{-1mm}

\begin{figure*}[t]
\includegraphics[width=\linewidth]{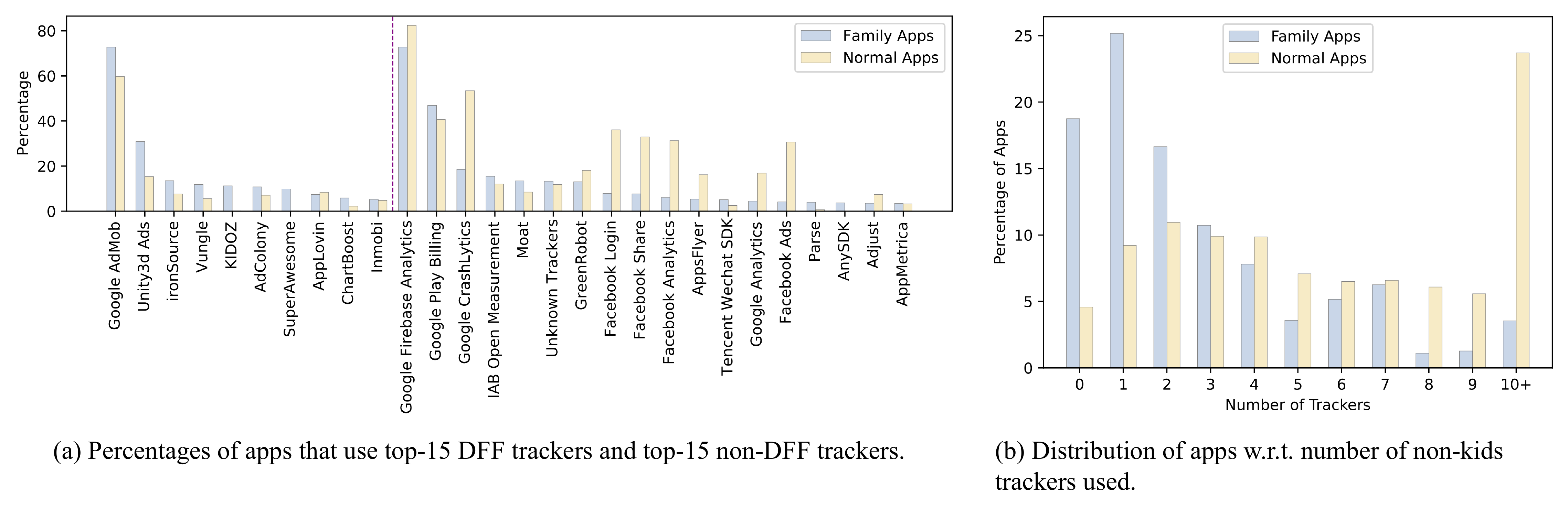}
\caption{Trackers analysis results.}
\label{fig_tracker}
\vspace{-3mm}
\end{figure*}

\subsection{Third-party Trackers}

To determine the usage and the potential privacy leakage through third-party trackers, we next leverage static and dynamic analysis. 

\noindent \textbf{The use of third-party trackers.~}
The static analysis results on the use of third-party trackers in both Family and Normal apps are presented in Figure~\ref{fig_tracker}(a). 
The trackers on the left side of the vertical red dash-line are allowed by the Design for Family program. 
\texttt{Google AdMob} is the most frequently used ad SDK in both Family and Normal apps. Not surprisingly, Family apps use the kids-allowed SDKs more frequently, especially for SDKs such as \texttt{KIDOZ}, \texttt{SuperAwesome}, and \texttt{ChartBoost}.
This could indicate that the Play store's Design for Family program contributes positively to app development --- more developers consider using the recommended SDKs when their target users are children. However, as shown by the right side of the vertical red dash-line, we find that quite a few (\trackersNotAllowedFamilyPerc) Family apps use SDKs that are not allowed for children's apps, \ie not in the Google Play certified ad SDKs.
For example, 72.8\% of Family apps use \texttt{Google Firebase Analytics} and the number for \texttt{Google Play Billing} is 46.91\%, while other tracker SDKs such as \texttt{Google CrashLytics}, \texttt{Facebook Login}, and \texttt{ApppsFlyer} are more frequently used in Normal apps. Even we exclude ads and tracker SDKs from Google, there are still \trackersNotAllowedFamilyPercExclude Family apps use at least one tracker that is not allowed for children's apps. 
Considering that some tracker SDKs are very popular among both Normal apps and Family apps, and they may provide irreplaceable functionalities to apps, we argue that a list of kids-allowed tracker SDKs is essential to protect the privacy of children users. This will likely also benefit app developers. 

Figure~\ref{fig_tracker}(b) shows the distribution of the number of trackers per-app.
Worryingly, only 18.75\% of Family apps do not use a disallowed ad or tracker SDK, whereas 95.43\% of Normal apps use SDKs that are disallowed when the target users include children. Even when we exclude SDKs from Google and Facebook, the numbers are still high: 38.49\% of Family apps and 28.25\% of Normal apps use at least one non-kids SDK.
In fact, 1.63\% of Family apps and 5.54\% of Normal apps use over 10 non-kids SDKs. From these results, we confirm that the Design for Family requirements are \emph{not} followed by all app developers. We note that, through static analysis, we cannot determine which ad or tracker SDKs are active during run-time. However, dynamic analysis may still be incomplete as it is hard to catch all tracking. We posit if the trackers are not necessary, the app developers should remove the SDKs from a children's app.

\noindent \textbf{Privacy leakage.~}
We further analyze the specific privacy leakages via both tracker SDKs and first-party APIs. 
We identify that 3.79\% of Family apps leak the device model, brand, builder, and other PII to third-party trackers without any consent from users. 
We observe that 7 Family apps sent \texttt{Private IP} data to \texttt{Google}, and \texttt{Supersonic Ads} obtains \texttt{Advertisement ID} (from 3 Family apps), \texttt{Location Information} (1 app), and \texttt{Android Serial} (from 3 Family  apps) without users' consents.
Furthermore, \texttt{Location Information} is sent to \texttt{Facebook} (from 4 Family apps), and \texttt{Babybus} collects the \texttt{Android Serial} (from 4 Family apps). We cannot check how the leaked PII is used, but emphasize that the leaking of PII has already violated GDPR and other privacy regulations (which is forbidden no matter if the target users are children or not). We note again that our results only present a lower-bound of the PII leakage without users' consent. 

\noindent \textbf{Case study: excessive tracking.~}
During the experiments, we found an interesting phenomenon: apps from the same developer tend to have similar trackers, even when the apps' functionalities or content are different. For example, we find that 6 apps from \texttt{Tu***ns}, a large developer in the children's game market (publishing over 100 games with 950+ million downloads according to its official website), has \textbf{47} trackers.
Although \texttt{Tu***ns} has a privacy policy which lists 17 ad or tracker SDKs, there are still 30 trackers that have not been mentioned. 
It is therefore unclear if parents or children know that their personal information could be shared with 17 (or even 47) third-party trackers.

\begin{figure}[t]
\centering
\includegraphics[width=\linewidth]{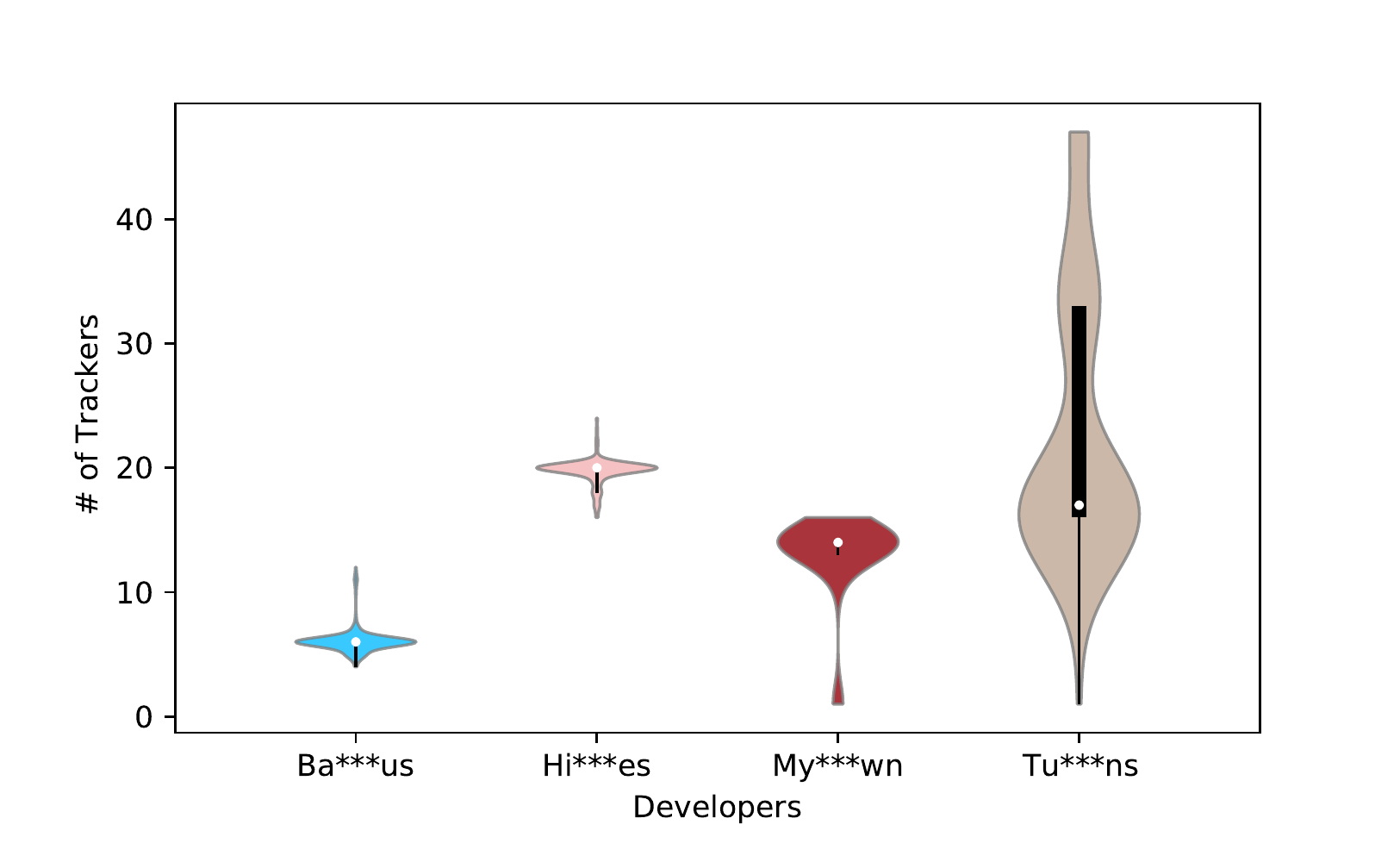}
\caption{Number of trackers used in children's apps from 4 big players.}
\label{fig_tracker_case}
\end{figure}

We investigate more apps to see if developers who have a large number (40+) of children's apps, embed a large numbers of trackers.
In Figure~\ref{fig_tracker_case}, we select four developers from our static results and list the distribution of numbers of trackers in their products. 
142 out of the 150 apps (94.7\%) from \texttt{Ba***us} have 5 to 6 trackers; 105 out of 133 apps (78.9\%) from \texttt{Hi***es} have 20 trackers, and 29 out of 43 apps (67.4\%) have 14 trackers. 
The distribution of the number of trackers is less concentrated among apps from \texttt{Tu***ns}: 6/62 \texttt{Tu***ns} apps have 47 trackers; 14 apps have 34 trackers, and 16 trackers are detected in the remaining 41 apps.
The reason behind this phenomenon could be that developers directly apply the same tracker configurations while developing different apps, instead of setting trackers for each app according to what personal information should be collected and shared.
Further, we conclude that most apps from these 4 big players contain trackers not allowed in children's apps.

\vspace{2mm}
\begin{mdframed}[backgroundcolor=white!10,rightline=true,leftline=true,topline=true,bottomline=true,roundcorner=2mm,everyline=true,nobreak=false] 
\noindent\textbf{Finding summary:}   
\begin{itemize} [leftmargin=*]
\item 81.25\% of Family apps use trackers that are not allowed for children; disallowed trackers are also frequently used in 95.43\% of Normal apps whose target user includes children. 
\item 1.63\% of Family apps and 5.54\% of Normal apps use over 10 ad or tracker SDKs that are not allowed to be used in children's apps.
\item Even big players do not follow the Play store policy. We find 4 developers who have 5--47 disallowed trackers in each of the 43 to 150 children's apps they listed on Play store.
\end{itemize}
\end{mdframed}
\vspace{-1mm}

\subsection{Inconsistent Content Age Ratings}
We next investigate the age ratings of the \appsWithRatings apps across all 5 rating authorities.
These are ACB for Australia, ESRB for Americas, PEGI for Europe and the Middle East, USK for Germany, and IARC (which is not country specific). 
We seek to check if the age ratings given by these different agencies are inconsistent.
We note that there could be demographic or cultural bias from parents. The purpose of releasing age ratings across different authorities is to provide a general reference message for parents' decision-making. Such information could be limited to only present highly inconsistent ratings.
Indeed, we find inconsistent ratings in \inconsistentAppsPerc apps, and \inconsistentAppsHighPerc apps have an inconsistency level above 3. This indicates that age rating inconsistencies among the various rating authorities are prevalent. 

\begin{figure}[t]
\centering
\includegraphics[width=\linewidth]{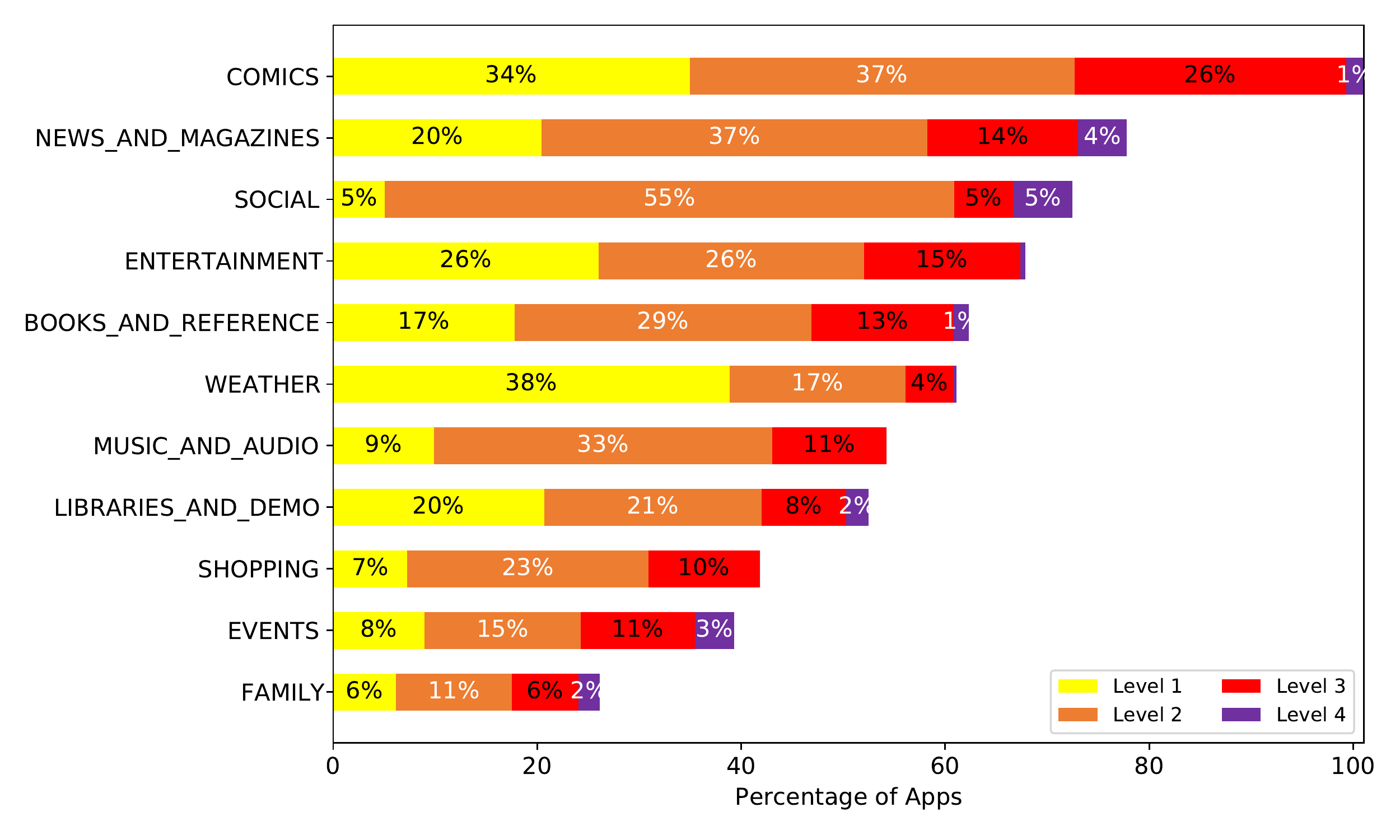}
\caption{Percentage of apps (per category) with inconsistent age ratings.}
\label{fig_rating_category}
\end{figure}

Figure~\ref{fig_rating_category} presents the age rating results across the top-10 categories of Normal apps \vs Family apps. Note, as one app may have multiple inconsistent rating pairs across different levels detected, the sum of apps could be higher than 100\%. 
Although Family apps have a lower ratio of inconsistent ratings than the other 10 Normal app categories, at least 6\% Family apps have an inconsistency level higher than 3. 26\% of \texttt{COMICS} and 15\% of \texttt{ENTERTAINMENT} apps have inconsistency levels above 3, and at least 55\% of \texttt{SOCIAL} apps have a rating inconsistency level 2. 
We conjecture that the reason for highly inconsistent ratings in these categories may be because the rules to flag sensitive content could be different across the rating authorities. However, we argue it is still not reasonable to have an app rated as 18+ in one territory and 3+ in another (level 4).
This will confuse parents and increase the risk to children (\eg allowing a child to play a game rated 18+ in another country). Therefore, we argue that, rather than only displaying the age ratings in the user's current territory, the app store should also provide ratings from other authorities as an option. Briefly listing the reason why the app is rated will also help parents and children to make a decision.

\noindent \textbf{Case study: highly inconsistent ratings.~}
Here, we showcase some inconsistent age ratings to highlight key concerns.
For example, \texttt{com.my***in.app}, 
a plant identification app with professional care guides (\num{1000000}+ installs), has been rated as ``PEGI 3'' and ``Rated for 3+'' in IARC. However, when we visit the Play store using \texttt{gl=de} and \texttt{gl=us}, the ratings change to ``USK 12+'' and ``ESRB Mature 17+''. 
The only explanation we can find is ``Drug use''. If the app truly contains information related to drugs, it is unsuitable for 3-year old children.
However, we manually check the app and only find drug-related information for plant diseases. Thus, the labels in PEGI 3 and IARC are unnecessary. 

Another example is \texttt{com.di***rd}, an instant messaging and digital distribution platform. As of 2021, the service has over 350 million registered users and over 150 million monthly active users (over \num{100}M installs from the Play store). As a social app, it is quite reasonable to be rated as ``Parental guidance'' in PEGI (as such apps may not always have predefined content that can be classified beforehand). 
It is rated as ``Teen'' in ESRB, yet in USK, the app is rated as ``18+'', which hints that the app contains ``drug use, realistic and explicit violence''. However, there is no explanation on the app page why the app is rated ``18+''.
We argue there should be better descriptions that explain why the app is rated, at least for apps not recommended for children.

\vspace{2mm}
\begin{mdframed}[backgroundcolor=white!10,rightline=true,leftline=true,topline=true,bottomline=true,roundcorner=2mm,everyline=true,nobreak=false] 
\noindent\textbf{Finding summary:}   
\begin{itemize}[leftmargin=*]
\item 21.69\% of apps have inconsistent content age ratings across different rating authorities.
\item There are many examples of confusing and inconsistent ratings, including among Family apps. 6\% of Family apps have an inconsistently level above 3.
\item The app store should provide ratings from other authorities as references for users, and explanations of ratings should be more transparent.\end{itemize}
\end{mdframed}
\vspace{-1mm}

\subsection{User Complaints}

Finally, we analyze the \numberComments comments downloaded from \appsWithComments apps in the Play store.
We strive to measure users' complaints across 4 categories: app content (where some content within is not suitable for children), advertisement, privacy, and security.  

Table~\ref{tab_comment_results} presents the percentage of comments that have been detected as complaints in each category, alongside the percentage of corresponding apps that they refer to.
In 2.12\% of the comments for Family apps, we recognize at least one user complaint relates to the app content, while the number is just 0.92\% for Normal apps.
This indicates that the users of Family apps complain more about the content. 
This reflects that the developers of Family app may not apply special controls on app content when the users are children. The ratios of comments related to advertisement complaints are similar in both Family and Normal apps, although this impacts more Normal apps than Family apps (34.81\% \vs 25.77\%). Further, fewer Family apps are complained about with respect to privacy and security than the other two categories. 
A possible explanation might be that non-technical users presumably cannot tell which trackers are embedded in an SDK, so they are unlikely to complain in an app comment about that particular privacy threat, but focus more on app content and advertisements in children's apps, which resonates with the contribution of our study. According to our measurements, 45 of the top-100 apps have quality rating scores above 4 stars and over \num{10}M installations. 

\begin{table}[t]
\caption{Comment analysis results.}
\label{tab_comment_results}
\begin{resizebox}{\linewidth}{!}{
\begin{tabular}{lcccc}
\toprule
\multirow{2}{*}{\textbf{Categories}} & \multicolumn{2}{c}{\textbf{Family Apps}} & \multicolumn{2}{c}{\textbf{Normal Apps}} \\ \cmidrule(r){2-3}\cmidrule(r){4-5}
 & \textbf{\% Comments} & \textbf{\% Apps} & \textbf{\% Comments} & \textbf{\% Apps} \\ \midrule
\textbf{Content} & 2.12\% & 30.62\% & 0.92\% & 33.86\%  \\
\textbf{Ads} & 1.01\% & 25.77\% & 0.97\% & 34.81\% \\
\textbf{Privacy} & 0.09\% & 6.91\% & 0.52\% & 29.05\% \\
\textbf{Security} & 0.16\% & 12.89\% & 0.43\% & 23.76\% \\ \bottomrule
\end{tabular}
}
\end{resizebox}
% \vspace{-6mm}
\end{table}

After we disclosed our findings to the Google privacy team, Play store updated the FAMILY category requirement, and explicitly required Teacher Approved badge for each kid-friendly app, where teachers and specialists rate apps based on criteria such as age and ads appropriateness, rather than self-certified~\cite{playstoreTeacherApproved}.
The Families policy has also been updated to declare that apps will be available only to users in regions where content within that app is deemed appropriate~\cite{playstoreLimitedRegions}. We provide more details of Play store policy updates in Appendix.

\vspace{2mm}
\begin{mdframed}[backgroundcolor=white!10,rightline=true,leftline=true,topline=true,bottomline=true,roundcorner=2mm,everyline=true,nobreak=false] 
\noindent\textbf{Finding summary:}  
\begin{itemize} [leftmargin=*]
\item 30.62\% and 25.77\% of Family apps have content-based and ad-based complaints, respectively. Users report privacy and security issues on Family apps less often though (6.91\% and 12.89\%, respectively).
\item Even highly popular and well rated apps accumulate many complaints --- among the top 100 apps that have the most complaint comments, 45 apps have app rating scores above 4 and over \num{10}M installs.
\end{itemize} 
\end{mdframed}

\section{Related Work}

In this section, we survey the recent research related to the evaluation of children's apps, static and dynamic analysis, and user comments analysis.

\noindent \textbf{Evaluation of children's apps.~}
Previous efforts have studied COPPA and children’s apps from various perspectives. Previous work examined the risks posed by third-party components bundled in children’s apps, with a focus on targeted advertisements~\cite{bhoraskar2014brahmastra,liu2016identifying}.
Several studies examined privacy policies~\cite{sun2020quality,massey2013automated,reyes2018won,reyes2017our} according to the requirements of regulations.
Other research has focused on methods aiding developers to make their apps more child-friendly in terms of content and privacy~\cite{liccardi2014can,hu2015protecting}.
Several works focus on the evaluation of app content age rating. For example, a study by Chen \etal~\cite{chen2013app} looks into the inappropriate age rating of mobile apps, exposing its potential risk for the children and adolescents. However, this research focuses on the comparison of age ratings between iOS app store and Google Play store, rather than exposing undesired application behaviors. 

\noindent \textbf{Static and dynamic analysis.~} 
Static code analysis techniques are widely used in the assessment of mobile apps~\cite{chen2020empirical,li2015iccta,xue2016you,chen2018mobile,liu2020maddroid,tang2020ios,wang2020clipboards,sun2021empirical,li2022cross,sun2022Measuring,hu2020mobile}. For example,  \textsc{FlowDroid}~\cite{arzt2014flowdroid} statically computes data flows in apps to understand which parts of the code that data may be exposed to. Notably, these off-the-shelf tools only utilize syntax-based scanning and data-flows, which leads to false positives that are not relevant to personal identifiable information. 
In contrast, dynamic analysis executes the code. Whereas static analysis often suffers from false positives, dynamic analysis is limited by the execution coverage~\cite{reardon201950}. Most work in this category focuses on analyzing apps’ network traffic~\cite{han2019you,le2015antmonitor,razaghpanah2018apps,shuba2018nomoads,van2017better}. In our work, we utilize static analysis for identifying permission requests and third-party trackers, while we trace privacy leakage using network traffic through a dynamic network monitoring tool~\cite{razaghpanah2018apps}.

\noindent \textbf{User comments analysis.~}
User comments of mobile apps have been extensively studied from several perspectives, including mining user opinions~\cite{chen2014ar,villarroel2016release,panichella2016ardoc,noei2019too,vu2016phrase}, app comment filtering~\cite{luiz2018feature,chen2014ar}, and exploring other concerns~\cite{nguyen2019short,chen2017toward}.  
Chen \etal~\cite{chen2017toward} conducted a study on
fraudulent campaigns to falsely boost apps’ rankings, which result in inappropriate risk exposure for
children and adolescents. 
Other work looks into research domains related to app descriptions, privacy policies, and other mobile app meta text, adopting NLP techniques~\cite{qu2014autocog}. 
PPChecker~\cite{yu2016can} identifies the inconsistencies between the sensitive behaviors of apps and their privacy policies. Recent research by Liu \etal~\cite{liu2021have} tries to solve the problem of compliance analysis between GDPR and privacy policies, utilizing a combination of sentence classification and rule-based analysis. However, the corpus suffers from an imbalanced data problem, which negatively affects the classification accuracy. We build on these techniques to investigates children's app behavior through comments analysis.

\section{Conclusion}
This paper has focused on the privacy practices of apps that are designed for children or have target users that include children. 
We have measured the use of permissions and trackers, investigated inconsistency in content age ratings, and analyzed user comment feedback. 
Our measurement results illustrate that, despite many privacy protection regulations and the strict requirements imposed by the app store, children still experience privacy threats.
This is caused by things like permission requests without child-friendly notifications, abuse of ad trackers,
confusing and inconsistent content age ratings, as well as privacy leakage without users consent. 
Ultimately, we conclude that the existing self-certification-based content age rating mechanism must be improved immediately.   

\section*{Ackowledgement}
The work has been supported by the Cyber Security Research Centre Limited whose activities are partially funded by the Australian Government’s Cooperative Research Centres Program. We thank Sai Teja Peddinti for useful discussion and guidance on the verification of our results.

\balance
\bibliographystyle{ACM-Reference-Format}
\bibliography{ref}

\section*{Appendix}
\setcounter{section}{0}

\section{Updates to Google Play Policies}
After we disclosed our findings to the Google privacy team, Play store announced several updates to Google Play Policies. Here we list some of the announcements and updates from \url{https://support.google.com/googleplay/android-developer/answer/9934569} that are related to our findings.

\begin{itemize}[leftmargin=*]
\item ``Effective May 11, 2022: We’re updating our Families policy to explain that if an app contains content that is not globally appropriate, we may make the app available only to users in regions where content within that app is deemed appropriate.''

\item Announced on July 27, 2022: ``We’re clarifying our Families Data Practices policy to state that apps that solely target children must not transmit Android advertising identifier (AAID), SIM Serial, Build Serial, BSSID, MAC, SSID, IMEI, and/or IMSI. Apps that target both children and older audiences must not transmit AAID, SIM Serial, Build Serial, BSSID, MAC, SSID, IMEI, and/or IMSI from children or users of unknown age.''

\item ``Effective November 1, 2022: We're updating our Families Self-Certified Ads SDK Program to require that self-certified ads SDK providers must submit new policy-compliant, self-certified versions and a test app to remain in the Families Self-Certified Ads SDK Program.''

\item Announced on November 16, 2022: ``To better align with the existing policy language, we're renaming the Designing Apps for Children and Families policy page to Google Play Families Policies. We're also consolidating the page formerly known as Families Ads \& Monetization onto the newly titled Google Play Families Policies page.''

\item ``Effective May 31, 2023: We’re updating our Families Self-Certified Ads SDK policy to require developers with apps in the Families Program to only use self-certified versions of SDKs when serving ads to children or users of unknown age.''
\end{itemize}

\end{document}